# ChatGPT-based Investment Portfolio Selection

**Oleksandr Romanko[1,3] · Akhilesh Narayan[2] · Roy H. Kwon[3]**

*First version: 11 August 2023*

**Abstract**  In this paper, we explore potential uses of generative AI models, such as ChatGPT, for investment portfolio selection. Trusting investment advice from Generative Pre-Trained Transformer (GPT) models is a challenge due to model "hallucinations", necessitating careful verification and validation of the output. Therefore, we take an alternative approach. We use ChatGPT to obtain a universe of stocks from S&P500 market index that are potentially attractive for investing. Subsequently, we compared various portfolio optimization strategies that utilized this AI-generated trading universe, evaluating those against quantitative portfolio optimization models as well as comparing to some of the popular investment funds. Our findings indicate that ChatGPT is effective in stock selection but may not perform as well in assigning optimal weights to stocks within the portfolio. But when stocks selection by ChatGPT is combined with established portfolio optimization models, we achieve even better results. By blending strengths of AI-generated stock selection with advanced quantitative optimization techniques, we observed the potential for more robust and favorable investment outcomes, suggesting a hybrid approach for more effective and reliable investment decision-making in the future.

**Keywords**  Portfolio optimization · Investment management · Generative AI · ChatGPT

## 1 Introduction

The advent of generative artificial intelligence (AI) models, particularly large language models (LLMs) like ChatGPT, marks a significant development and can potentially disrupt different industries, including finance [1,10,11,13,17]. These models, based on the transformative architecture of generative AI, can generate detailed, contextually coherent outputs by analyzing massive text datasets, and have demonstrated impressive capabilities in tasks ranging from creative writing to complex problem-solving. The use of these models, ChatGPT in particular, for selection of investment strategies or "stock picking" has drawn significant attention [2,3,4,8,9,14] since the launch of ChatGPT in November 2022 and especially GPT-4 [15], its more advanced successor, in mid-March 2023. Other LLMs available at the time of writing this paper, such as Bard from Google or Claude 2 from Anthropic, may be considered as alternatives to GPT models from OpenAI.

However, the question of how effective ChatGPT might be in selecting financial assets, such as stocks, for investing, remains open. One challenge arises from the black-box nature of generative AI models, which involve complex non-linearities, making it difficult to discern how they analyze their training data to define investment strategies. The specifics of the training data used for the GPT-4 model are unknown, adding

Oleksandr Romanko
oleksandr.romanko@sscinc.com

Akhilesh Narayan
akhilesh.narayan@iitb.ac.in

✉ Roy H. Kwon
rkown@mie.utoronto.ca

1 SS&C Algorithmics, 200 Front Street West, suite 2500, Toronto, ON, M5V3K2, Canada

2 Department of Mechanical Engineering, Indian Institute of Technology Bombay, Mumbai, 400076, India

3 Department of Mechanical and Industrial Engineering, University of Toronto, 5 King's College Road, Toronto, ON, M5S3G9, Canada

another layer of uncertainty. Furthermore, GPT models are subject to 'hallucinations'. This refers to instances where the AI, although generating text based on its training, does so without a solid conceptual framework behind it. This issue is particularly problematic for investment strategies, as erroneous or misleading information could lead to significant financial repercussions, underscoring the need for rigorous validation and verification before these models can be utilized in practice.

Despite mentioned challenges, the potential of these models remains significant. By rigorously testing, validating, and refining the outputs of ChatGPT, i.e., GPT-selected portfolios, with proven quantitative finance models like portfolio optimization, we may unlock new opportunities in the realm of investment strategies selection.

In this study, we utilized the GPT-4 model through its API. At the time of writing, the GPT-4 model was trained on data prior to September 2021. Consequently, we used historical data up prior to September 2021 as our in-sample data for portfolio selection models and post-September 2021 data was used for out-of-sample testing.

As the GPT-4 model was trained on large text datasets, it should be able to indirectly derive "sentiment" towards performance and even risks of various stocks (and potentially other financial assets like bonds). In addition to biases incorporated in datasets used for training, e.g., some stocks may be mentioned much more frequently in text data, there are biases related to probabilistic nature of generative AI. Hence, ChatGPT may not be able to "reason" well about why it picks certain stocks. The underlying hypothesis is that the frequency and positivity of mentions related to a given stock in its training data (by different analysts, in the company reports, in blog posts, news articles, research publications and other text documents) may influence its selection. Thus, stocks that received significant "positive sentiment" or were associated with successful investment and trading strategies in the training datasets are more likely to be recommended by ChatGPT. It is not clear if risks are accounted at the same level as performance/reward.

In this study, we aim to leverage a vast dataset on which ChatGPT was trained to identify "most popular" or "most performing" stocks. We then test stocks suggested by ChatGPT and simple strategies based on those and apply quantitative finance techniques, such as portfolio optimization, to enhance the suggested strategies. To conduct our experiments, we asked ChatGPT to generate three distinct trading universes of stocks from the S&P500 market index, with the goal of outperforming the index. These trading universes varied in size, consisting of 15, 30, and 45 stocks, respectively. We further asked ChatGPT to assign asset weights to each of the stocks, leading to the creation of three distinct ChatGPT-weighted portfolios (one for each trading universe).

However, an inherent challenge with this approach is that ChatGPT may return different asset universes each time we send a request to the GPT-4 API. Hence, to increase robustness, we repeated the API request to ChatGPT 30 times and kept a record of how frequently each stock was found in the output. From these results, we selected the 15, 30, and 45 most frequently suggested stocks for our respective portfolios. In some rare cases, it may also return to us stocks that are not in the S&P500 index. To address this issue, we verified whether each of the assets returned by ChatGPT is in the S&P500 market index.

A second goal of this study is to compare portfolios suggested by the GPT-4 model with those derived using established portfolio optimization techniques. Specifically, we computed the Markowitz mean-variance efficient frontier [12] for the universe of assets suggested by ChatGPT. To ensure a fair comparison, we imposed constraints on asset weights within our optimized portfolios. More precisely, we mandated that each asset's weight should be approximately between half and double the equal weight of $(1/n)$, where $n$ is the number of assets in the portfolio. For instance, with a portfolio of 15 assets, the weight of each asset would be confined to the range between 3% and 13%. Simultaneously, the sum of all weights was constrained to equal one. Subsequently, using five years of weekly data prior to September 2021, we calculated a vector of expected returns and covariance matrix of returns. On this in-sample data, we perform mean-variance optimization and plot the Markowitz efficient frontier. From the efficient frontier, we selected three key portfolios: minimum variance portfolio, maximum expected return portfolio, and maximum Sharpe ratio portfolio [16]. We conducted this process for trading universes of size 15, 30, and 45, resulting in five distinct portfolios of each size for comparison, namely GPT-weighted portfolio, equally weighted portfolio, minimum variance portfolio, maximum expected return portfolio, and maximum Sharpe ratio portfolio.

In addition to this, we decided to include portfolios created by cardinality-constrained optimization from the S&P500 universe of stocks. This involved running cardinality-constrained optimization problems [5], computing Markowitz efficient frontier with an additional constraint ensuring a selection of exactly 15 out of all S&P500 stocks, and then repeating the process for 30 and 45 stocks. To maintain fairness, i.e., comparing portfolios of the same size, we also placed minimum and maximum holding constraints on asset weights, ensuring that the weight of each stock either fell between the lower and upper bounds or was set to zero in the course of optimization. After calculating these cardinality-constrained efficient frontiers, we selected three portfolios from each – minimum variance, maximum expected return, and maximum Sharpe ratio portfolios – to add to our comparison. The rationale behind using cardinality-constrained optimization was to thoroughly evaluate ChatGPT's "stock picking" capability. By exploring all possible combinations of stocks within the S&P500 universe, we could assess the model's ability to select promising investment options, ensuring a comprehensive and rigorous analysis of portfolio construction strategies.

Finally, we evaluated and compared performance and risk measures of different sets of portfolios mentioned above – each set containing eight portfolios of sizes 15, 30, and 45 respectively – comparing those against each other and the S&P500 index. In addition to S&P500 index, we also compare to the Dow Jones Industrial Average and Nasdaq Composite indices. For a more comprehensive comparison, we also benchmark all portfolios against well-known investment funds from asset management companies such as Vanguard, Fidelity, and Blackrock. These popular funds are widely acknowledged and used, reflecting a broader market performance in the United States and globally. The portfolios were assessed on a variety of metrics, including returns, volatility, Sharpe ratio, maximum drawdown and value-at-risk among others. For out-of-sample testing, we used three distinct time periods. The first period was chosen to be from September 2021 to July 2023, as our in-sample data was prior to September 2021. The remaining two shorter periods, from March 14, 2023, to July 2023 and May 2023 to July 2023, were chosen to account for the potential impacts of the release of GPT-4 and the actual time at which the portfolios were selected using GPT-4 prompts. A comprehensive discussion on the comparison of these portfolios, their performances, and implications are presented in the results section of the paper.

## 2 GPT-4 Selection of Trading Universe of Stocks

To obtain trading universes of assets, we used the GPT-4 model by providing the following prompt: "Using a range of investing principles taken from leading funds, create a theoretical fund comprising of at least X stocks (mention their tickers) from the S&P500 with the goal to outperform the S&P500 index", (where 'X' represents 15, 30, or 45 assets). We leveraged the GPT-4 API automation and sent each request 30 times to capture a broad range of responses. The GPT-4 API was used to extract stock tickers included in each response. After each iteration, we updated a dictionary consisting of the stock tickers and their frequency of appearances in the GPT-4 output response. It is also worth noting here that when the above prompt is used without the phrase – "comprising of at least X stocks", the outputs are portfolios consisting of about 15 stocks on average. That is the reason why we generate portfolios of 15 stocks and its multiples 30 and 45. The latter were considered to generate more diversified portfolios.

When asked to generate a portfolio of at least 15 stocks, GPT-4 produced between 15 to 20 stocks. Similarly, the request for 30 and 45 stocks resulted in GPT-4 providing 30 to 35 stocks and 45 to 50 stocks, respectively. Stocks that appeared most frequently in GPT-4's responses were subsequently filtered and selected for our trading universe. Figures 1, 2, and 3 illustrate the final selection of the trading universes of 15, 30, and 45 stocks, respectively. The most frequently mentioned stocks are highlighted in red, while other stocks appearing at least once in GPT-4's responses are marked in blue.

Note that we had to replace ticker "FB" obtained from GPT-4 with "META". As the GPT-4 model was trained on data prior to September 2021, it was unaware of the renaming of "FB" to "META" that occurred in 2022.

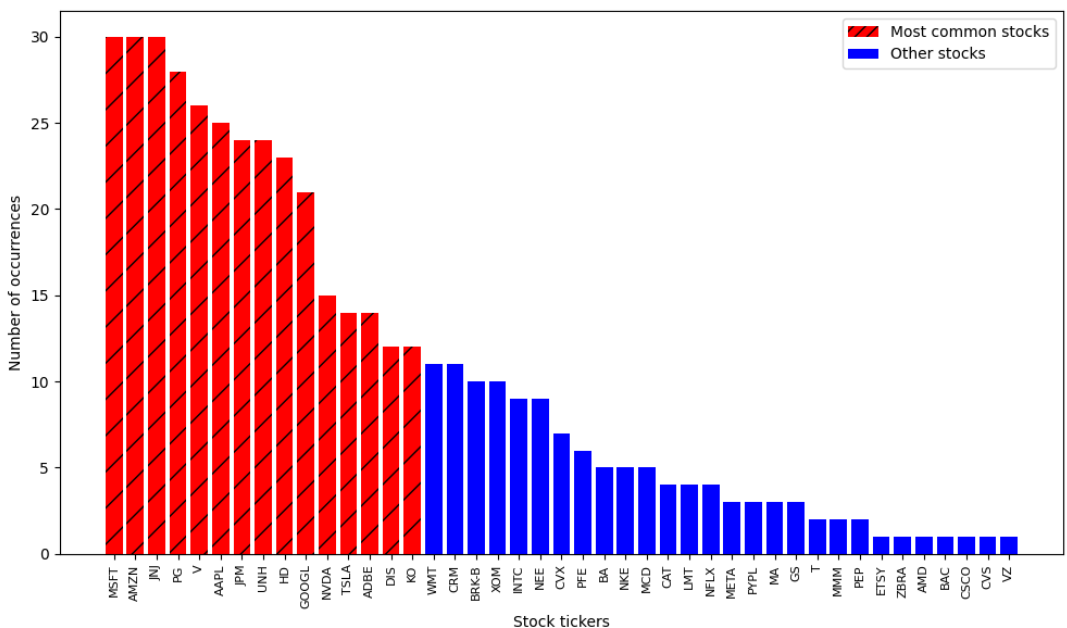

**Fig. 1** Universe of 15 stocks selected by GPT-4

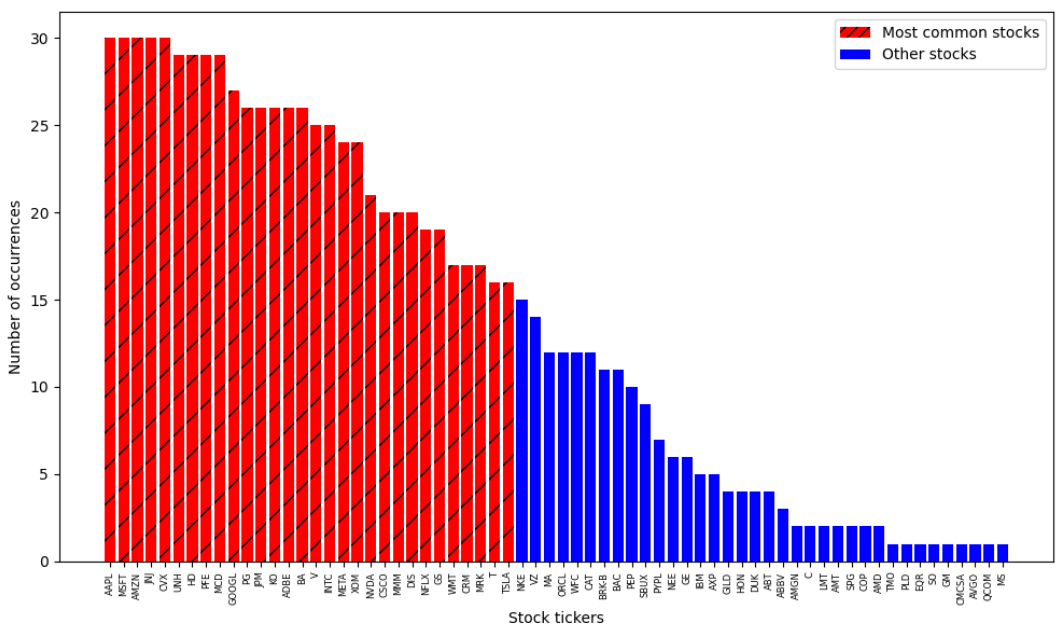

**Fig. 2** Universe of 30 stocks selected by GPT-4

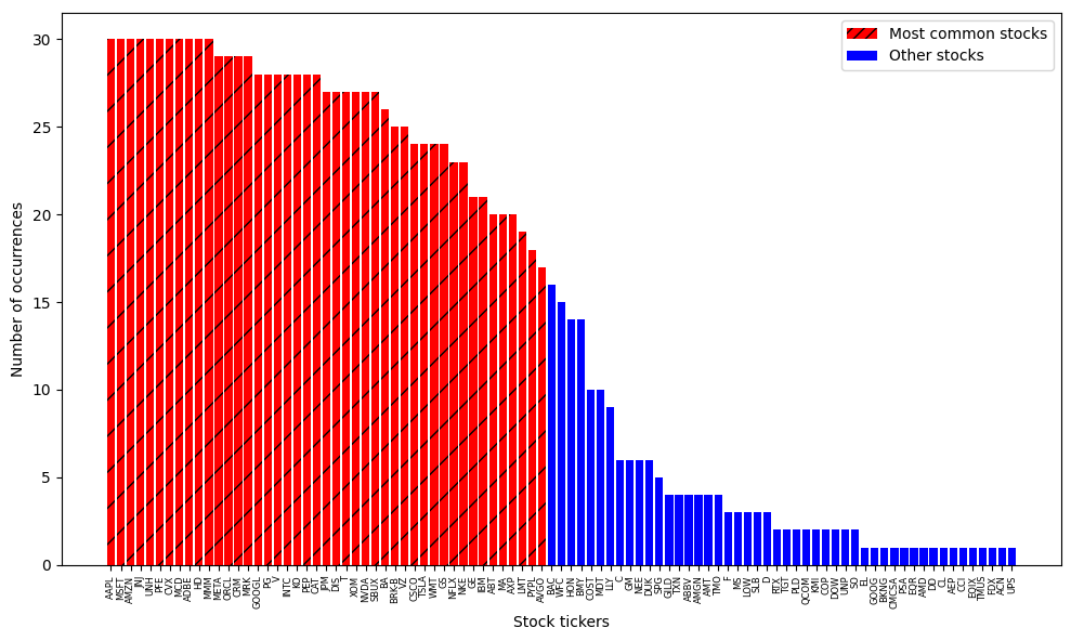

**Fig. 3** Universe of 45 stocks selected by GPT-4

In Table 1 we analyzed overlaps between stocks selected for the universe of 15, 30 and 45 assets. It also shows the weights of each stock in the "GPT-weighted" portfolios for each of the three universes.

| ticker | MSFT | KO | DIS | ADBE | TSLA | GOOGL | HD | NVDA | JPM | AAPL | V | PG | JNJ | AMZN | UNH |
|---|---|---|---|---|---|---|---|---|---|---|---|---|---|---|---|
| 15 Stocks Universe | 9.17% | 4.59% | 4.59% | 4.59% | 4.59% | 7.34% | 6.42% | 7.34% | 6.42% | 9.17% | 7.34% | 6.42% | 6.42% | 9.17% | 6.42% |
| 30 Stocks Universe | 5.04% | 2.88% | 2.88% | 4.32% | 3.6% | 4.32% | 2.88% | 3.6% | 3.6% | 5.04% | 3.6% | 2.88% | 3.6% | 4.32% | 3.6% |
| 45 Stocks Universe | 4.08% | 2.04% | 2.04% | 2.72% | 2.72% | 2.72% | 2.04% | 2.72% | 2.04% | 4.08% | 2.72% | 2.04% | 2.72% | 4.08% | 2.72% |

| ticker | NFLX | T | MRK | CRM | WMT | GS | MMM | CSCO | XOM | META | INTC | BA | MCD | PFE | CVX |
|---|---|---|---|---|---|---|---|---|---|---|---|---|---|---|---|
| 15 Stocks Universe | | | | | | | | | | | | | | | |
| 30 Stocks Universe | 2.88% | 2.88% | 2.88% | 2.88% | 2.88% | 2.88% | 2.88% | 2.88% | 2.88% | 3.60% | 2.88% | 2.88% | 2.88% | 2.88% | 2.88% |
| 45 Stocks Universe | 2.04% | 1.36% | 2.04% | 2.04% | 2.04% | 2.04% | 2.04% | 2.04% | 2.04% | 2.04% | 2.04% | 2.04% | 2.04% | 2.04% | 2.04% |

| ticker | LMT | AXP | MA | ABT | IBM | GE | SBUX | VZ | BRK-B | PYPL | CAT | PEP | ORCL | NKE | AVGO |
|---|---|---|---|---|---|---|---|---|---|---|---|---|---|---|---|
| 15 Stocks Universe | | | | | | | | | | | | | | | |
| 30 Stocks Universe | | | | | | | | | | | | | | | |
| 45 Stocks Universe | 2.04% | 2.04% | 2.04% | 2.04% | 1.36% | 1.36% | 2.04% | 1.36% | 2.04% | 2.04% | 2.04% | 2.04% | 2.04% | 2.04% | 2.04% |

**Table 1** Overlaps between assets in three universes and asset weights assigned by ChatGPT

From each of the three trading universes, we initially constructed two distinct portfolio types. Firstly, we created an "equally-weighted" ($1/n$) portfolio, wherein the weight attributed to each asset was calculated as the reciprocal of the total number of stocks in that universe (15, 30, or 45). Secondly, we solicited weight assignments for these assets from ChatGPT using the following GPT-4 prompt: "Assume you're designing a theoretical model portfolio from these S&P500 stocks: {input}. Provide a hypothetical example of how you might distribute the weightage of these stocks (normalized, i.e., weights should add up to 1.00) in the portfolio to potentially outperform the S&P500 index. Also mention the underlying strategy or logic which you used to assign these weights." Here, the {input} comprised of the list of X most frequently occurring stock tickers as per the earlier prompt.

ChatGPT rationalized its weight assignments by citing the consideration of various factors such as sector diversification, market capitalization, growth potential, and stability among others (see Appendix B for more details about GPT-4 prompts that we used and obtained responses). This second portfolio is referred to as the "GPT-weighted" portfolio. Weights of "GPT-weighted" portfolios are shown in Table 1.

Figure 4 shows sector weights for the "GPT-weighted" portfolios of 15, 30 and 45 stocks.

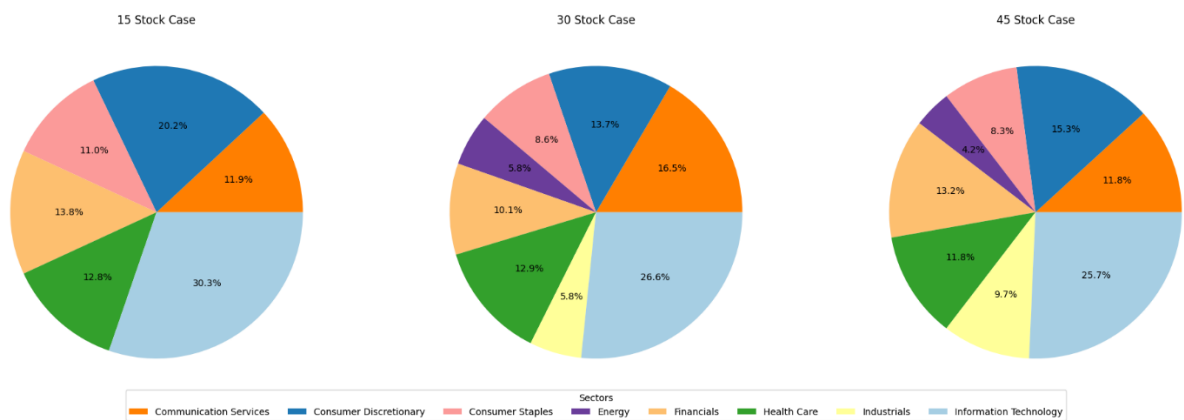

**Fig. 4** Allocation of weights to sectors for "GPT-weighted" portfolio of 15, 30 and 45 stocks

As the portfolio size increases from 15 to 45 stocks, there is a noticeable trend towards greater diversification across various sectors. Despite "Information Technology" remaining dominant, its proportion decreases, allowing sectors like "Industrials", "Energy", and "Communication Services" to play a more significant role, thereby potentially mitigating risk and capturing wider market opportunities.

By now for each of the three asset universes, we obtained two portfolios – "equally-weighted" and "GPT-weighted", where weights of each stock are equal or assigned by ChatGPT. In the next section, we compute additional portfolios for each universe based on portfolio optimization algorithms.

## 3 Portfolio Optimization Models and Computational Tests

### 3.1 Data

To compare reward and risk measures across the three trading universes, in addition to the two portfolios described in the previous section, we also computed optimal portfolios on the efficient frontier according to the Markowitz mean-variance portfolio selection model with additional constraints. Before we describe our computations in more detail, we will discuss data for running our optimization problems and estimation of expected returns and covariances. To align with the GPT-4 training data period, we utilized historical financial data up until September 2021 for our optimization formulations, subsequently evaluating portfolio performance using out-of-sample data from September 2021 to July 2023.

In summary, we used the following data:
- Weekly data for five years prior to September 2021, utilized for expected return and covariance matrix computations (in-sample data for optimization). As GPT-4 model was trained on data prior to September 2021, we decided to use historical data prior to September 2021 for calculating parameters for the mean-variance portfolio selection model and its extensions.

- Out-of-sample data used for testing and comparing portfolios span three distinct periods:
    1) Out-of-sample period 1: from September 2021 to July 2023, representing the total post-training data period of GPT-4;
    2) Out-of-sample period 2: from March 14, 2023 (the release date of GPT-4) to July 2023, to gauge performance post the model's release;
    3) Out-of-sample period 3: from May 2023, when we asked ChatGPT to generate the portfolios, to July 2023.

From all stocks included in the S&P500 market index, we obtained complete historical timeseries (both in-sample and out-of-sample) for 485 stocks using Yahoo Finance. As a result, we restricted our attention to only those 485 assets.

### 3.2 Mean-Variance Portfolio Optimization Model

By now, ChatGPT has selected three trading universes: 15, 30 and 45 stocks from S&P500. We already have "equally-weighted" and "GPT-weighted" portfolios for each of the three universes. We further computed the mean-variance efficient frontier for each universe and selected three additional portfolios:

- Minimum variance portfolio ("min Var");
- Maximum expected return portfolio ("max Ret");
- Maximum Sharpe ratio portfolio ("max Sharpe").

We use a set of standard constraints for our portfolio construction process which include the asset weights summing up to 100% and bound constraints on asset weights:

$$\Sigma_{i=1}^{n} w_i = 1$$
$$l \leq \boldsymbol{w} \leq u \quad (1)$$

where $n$ denotes the number of assets in a universe and $\boldsymbol{w}$ is the vector of (unknown) asset weights. To ensure a fair comparison of all portfolios, we have added bound constraints $l \leq \boldsymbol{w} \leq u$. Bound constraints ensure that each coordinate of vector $\boldsymbol{w}$ is between lower bound $l$ and upper bound $u$. Typical bound constraints are no short sale constraints, where $l = 0$ and $u = 1$. In our case, lower bound $l$ is set to be approximately half of the weight in the equally-weighted portfolio, and upper bound $u$ is set to approximately twice of the weight in the equally-weighted portfolio. In summary, the bound constraints are:

- $0.03 \leq \boldsymbol{w} \leq 0.13$ for the universe of 15 stocks;
- $0.02 \leq \boldsymbol{w} \leq 0.07$ for the universe of 30 stocks;
- $0.01 \leq \boldsymbol{w} \leq 0.05$ for the universe of 45 stocks.

Bound constraints ensure that all assets have non-zero weights. As a result, all portfolios that we compare will have the same number of assets (15, 30 or 45).

For the stocks selected by GPT-4, the portfolio optimization problem for computing mean-variance efficient frontier subject to constraints (1) can be defined as:

$$\begin{aligned} \min_{\boldsymbol{w}} \quad & \boldsymbol{w}^T \cdot \boldsymbol{Q} \cdot \boldsymbol{w} \\ \text{s.t.} \quad & \boldsymbol{\mu}^T \cdot \boldsymbol{w} \geq \varepsilon \\ & \Sigma_{i=1}^{n} w_i = 1 \\ & l \leq \boldsymbol{w} \leq u \end{aligned} \quad (2)$$

where $\boldsymbol{\mu}$ is the vector of expected returns, and $\boldsymbol{Q}$ is the covariance matrix of returns. Changing hyperparameter $\varepsilon$ (target expected return) allows us to compute mean-variance efficient frontier, where each portfolio on the frontier contains exactly 15, 30 or 45 assets.

Figure 5 shows an efficient frontier of 15 assets on in-sample data. For comparison, dashed line shows efficient frontier where bound constraints $l \leq w \leq u$ in (2) are replaced with no-short-sale constraints $w \geq 0$. Figure 6 and Figure 7 show efficient frontiers of 30 and 45 asset universes on the in-sample data.

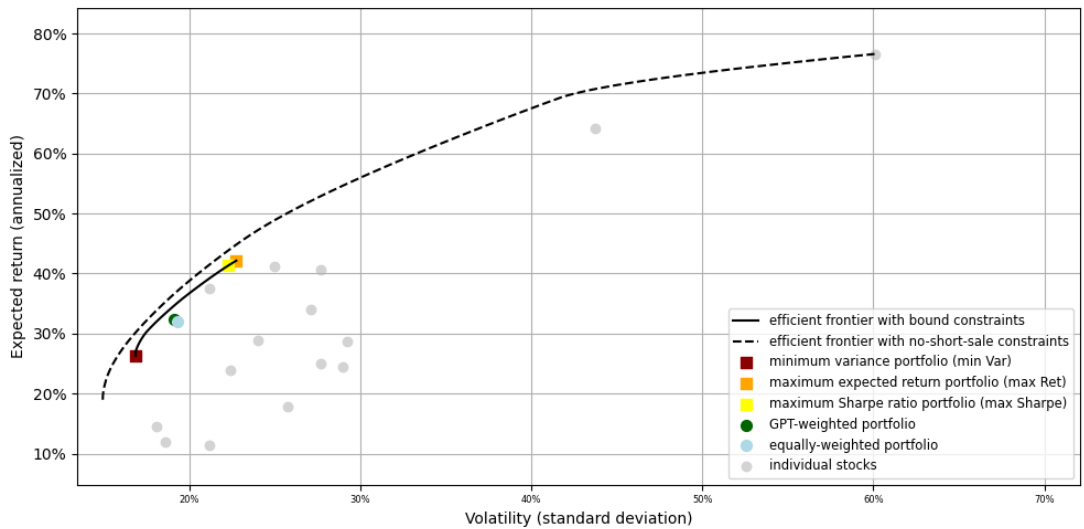

**Fig. 5** Risk-reward profiles of portfolios from the universe of 15 stocks (in-sample)

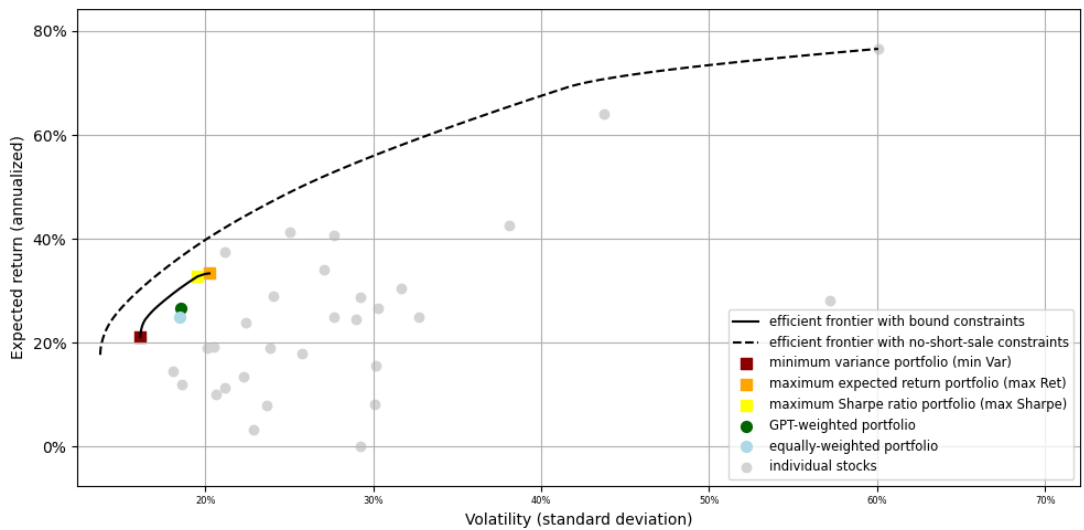

**Fig. 6** Risk-reward profiles of portfolios from the universe of 30 stocks (in-sample)

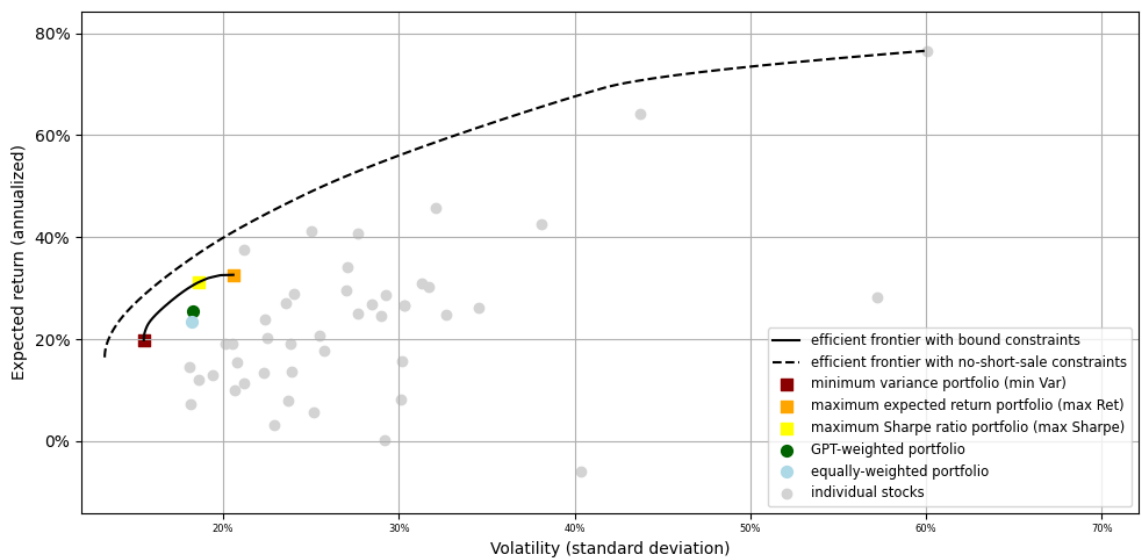

**Fig. 7** Risk-reward profiles of portfolios from the universe of 45 stocks (in-sample)

Our observations indicate that both the "GPT-weighted" and "equally-weighted" portfolios are positioned near the efficient frontier and are approximately equidistant to maximum expected return and minimum variance portfolios. Notably, the "GPT-weighted" portfolio tends to be slightly closer to the efficient frontier as compared to the "equally-weighted" portfolio across all three cases.

### 3.3 Mean-Variance Cardinality-Constrained Portfolio Optimization Model

We considered cardinality constraints in our analysis, as those ensure that a portfolio holds a specific number of assets. This approach allowed us to thoroughly evaluate ChatGPT's "stock picking" capability by exploring all possible combinations of stocks within the S&P500 universe.

In addition to the already computed portfolios for each universe – the "equally-weighted", "GPT-weighted", "min Var", "max Ret", and "max Sharpe" – we also computed mean-variance efficient frontiers incorporating cardinality constraints and minimum/maximum holding constraints. Cardinality constraint in portfolio optimization ensures that a portfolio holds a specific number of assets, and for our purposes, we have set cardinality at three levels: 15, 30, and 45, out of the available 485 assets included in the S&P500 index. With each cardinality level ($K = 15, 30, 45$), we compute the efficient frontier, and then select three additional portfolios on it:

- Minimum variance cardinality-constrained portfolio ("min Var card");
- Maximum expected return cardinality-constrained portfolio ("max Ret card");
- Maximum Sharpe ratio cardinality-constrained portfolio ("max Sharpe card").

Using the convention $0^0 = 0$, the cardinality function of a vector $\boldsymbol{w}$ is defined as $\text{card}(\boldsymbol{w}) = \Sigma_{i=1}^{n} w_i^0$. In the investment portfolio management context of this study, the cardinality function is interpreted as the number of assets held in the portfolio, with non-zero individual weights.

To restrict the number of assets held in a portfolio to be equal to $K$, we add the following cardinality constraint to our optimization problem formulation (2):

$$\begin{aligned} \min_{\mathbf{w},\mathbf{z}} \quad & \mathbf{w}^T \cdot \mathbf{Q} \cdot \mathbf{w} \\ \text{s.t.} \quad & \boldsymbol{\mu}^T \cdot \mathbf{w} \geq \varepsilon \\ & \Sigma_{i=1}^{n} w_i = 1 \\ & 0 \leq \mathbf{w} \leq u \\ & \Sigma_{i=1}^{n} z_i = K \\ & l \cdot \mathbf{z} \leq \mathbf{w} \\ & \mathbf{z} \in \{0,1\} \end{aligned} \quad (3)$$

Cardinality is a counting function which requires the introduction of binary variables $z_i$ into the problem formulation. The resulting optimization problem is known as mixed-integer optimization problem (MIP). Efficient frontiers computed by varying the hyperparameter $\varepsilon$ include portfolios with cardinality $K$, where the weight of each asset is between the bounds $l \leq \mathbf{w} \leq u$.

We set cardinality $K = 15, 30, 45$ and compute three efficient frontiers for cardinality constrained optimization problem (3). All optimization computations are performed in Python with CVXPY module [6] and CPLEX solver [7]. It took 243, 81 and 70 seconds respectively for the CPLEX solver to compute efficient frontiers of 100 portfolios for cardinality 15, 30 and 45.

Cardinality-constrained efficient frontiers for the three universes with and without minimum/maximum holding constraints on asset weights are shown in Figure 8. Efficient frontiers shown with solid lines are computed by optimization formulation (3), while efficient frontiers shown by dashed curves are computed by optimization formulation (2).

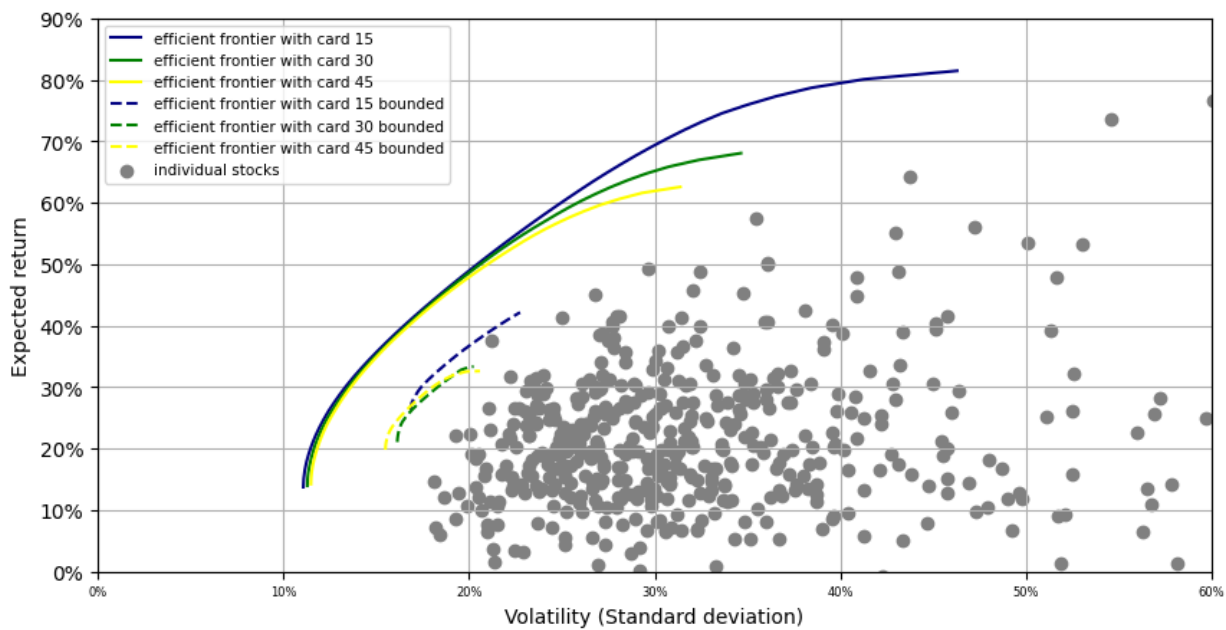

**Fig. 8** Risk-reward profiles of portfolios with cardinality 15, 30 and 45 (in-sample)

As expected, we observe that the efficient frontiers with bounded asset weights computed by optimization formulation (2) are located beneath their cardinality-constrained counterparts computed by the formulation (3). It is due to cardinality-constrained optimization computing optimal portfolios from the wider trading universe of 485 stocks, while the formulation (2) operates on reduced trading universes of 15, 30 and 45 stocks selected by ChatGPT. Performance of these cardinality-constrained portfolios is assessed in the next section.

### 3.4 Comparison

To ensure a fair comparison, in the next section of the paper we will examine out-of-sample performances of eight distinct portfolios, each composed of 15, 30, or 45 stocks:

- Strategies type 1:
    1. portfolio from GPT trading universe with weights assigned by GPT-4 model ("GPT-weighted");
    2. portfolio from GPT trading universe with equal weights ("equally-weighted");
- Strategies type 2:
    3. minimum variance portfolio from GPT trading universe ("min Var");
    4. maximum expected return portfolio from GPT trading universe ("max Ret");
    5. maximum Sharpe ratio portfolio from GPT trading universe ("max Sharpe");
- Strategies type 3:
    6. minimum variance cardinality-constrained portfolio ("min Var card");
    7. maximum expected return cardinality-constrained portfolio ("max Ret card");
    8. maximum Sharpe ratio cardinality-constrained portfolio ("max Sharpe card").

For out-of-sample comparison, in addition to S&P500 market index, we also compare with the Dow Jones Industrial Average and the Nasdaq Composite indices.

In our effort to perform a comprehensive comparison we also compare with well-known investment funds to capture the overall market's performance. We take an average of the performance (cumulative returns) of these funds in the out-of-sample periods and represent it by a "Popular Investment Funds" portfolio:

- Vanguard Total World Stock ETF (VT);
- Fidelity Large Cap Growth Index (FSPGX);
- Vanguard Total Stock Market ETF (VTI);
- Fidelity SAI U.S. Quality Index Fund (FUQIX);
- Vanguard Mid-Cap Value Index Fund (VMVAX);
- Fidelity Contrafund (FCNTX);
- Vanguard Global Equity Fund (VHGEX);
- Fidelity Worldwide Fund (FWWFX);
- T. Rowe Price Global Stock Fund (PRGSX);
- Blackrock Advantage Global Fund (MDGCX);
- JPMorgan Large Cap Growth Fund (OLGAX);
- Fidelity Growth and Income (FGIKX);
- Vanguard Dividend Growth (VDIGX).

## 4 Results

While evaluating and comparing our results, we can conclude from the out-of-sample cumulative return comparison that the portfolios constructed from the ChatGPT trading universes outperform those derived from cardinality-constrained optimizations, implying that ChatGPT is good at stock pre-selection. This applies across all strategies with asset universes of size 15, 30, and 45. Figures 9, 10, and 11 show cumulative returns for the 15 stock portfolios and benchmarks during out-of-sample period 1, period 2, and period 3, respectively. In Tables 2, 3, and 4, we compare performance and risk metrics of various portfolios within the 15 stock universe for the corresponding out-of-sample periods. The figures and tables for the 30 and 45 stock universes are shown in Appendix A.

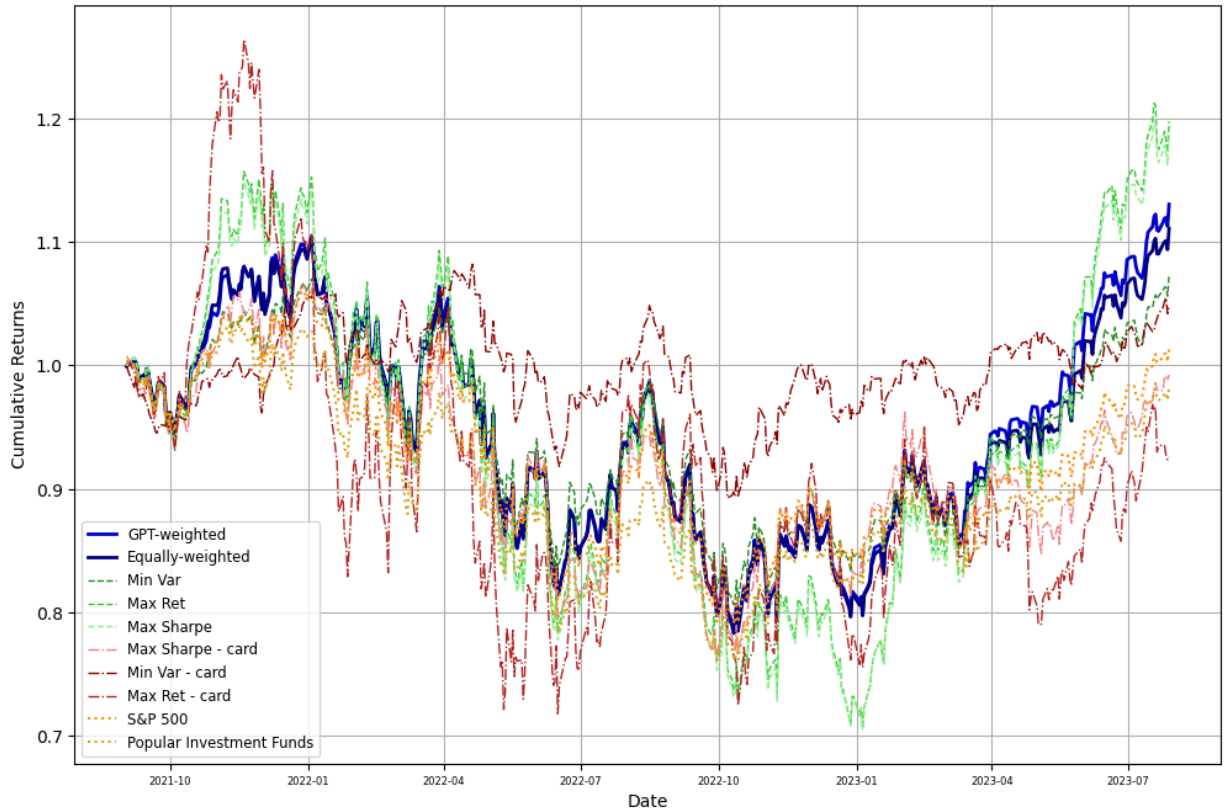

**Fig. 9** Cumulative returns of portfolios with 15 assets and benchmarks out-of-sample (1 September 2021 – 31 July 2023)

| | Cumulative Returns | Expected Return | Volatility of Return | Max Drawdown | Sharpe Ratio | VaR 99% of Return |
|---|---|---|---|---|---|---|
| GPT-weighted | 113.04% | 0.17% | 3.09% | -27.75% | 0.39 | -3.92% |
| Equally-weighted | 111.09% | 0.15% | 3.07% | -27.73% | 0.36 | -3.83% |
| Min Var | 107.32% | 0.11% | 2.69% | -23.68% | 0.29 | -3.29% |
| Max Ret | 119.76% | 0.26% | 4.02% | -37.53% | 0.47 | -4.63% |
| Max Sharpe | 118.68% | 0.25% | 3.97% | -37.09% | 0.45 | -4.56% |
| Max Sharpe - card | 99.24% | 0.03% | 3.17% | -27.77% | 0.08 | -3.73% |
| Min Var - card | 104.88% | 0.07% | 2.17% | -17.09% | 0.25 | -2.47% |
| Max Ret - card | 92.59% | 0.05% | 5.11% | -42.64% | 0.09 | -5.76% |
| S&P 500 | 101.29% | 0.04% | 2.62% | -24.82% | 0.13 | -3.42% |
| Dow Jones | 100.42% | 0.03% | 2.32% | -20.95% | 0.09 | -2.80% |
| NASDAQ | 93.52% | -0.01% | 3.31% | -35.72% | -0.00 | -4.20% |
| Popular Investment Funds | 98.25% | 0.01% | 2.55% | -26.00% | 0.04 | -3.32% |

**Table 2** Evaluation metrics (weekly) for portfolios with 15 stocks and benchmarks from 1 September 2021 to 31 July 2023

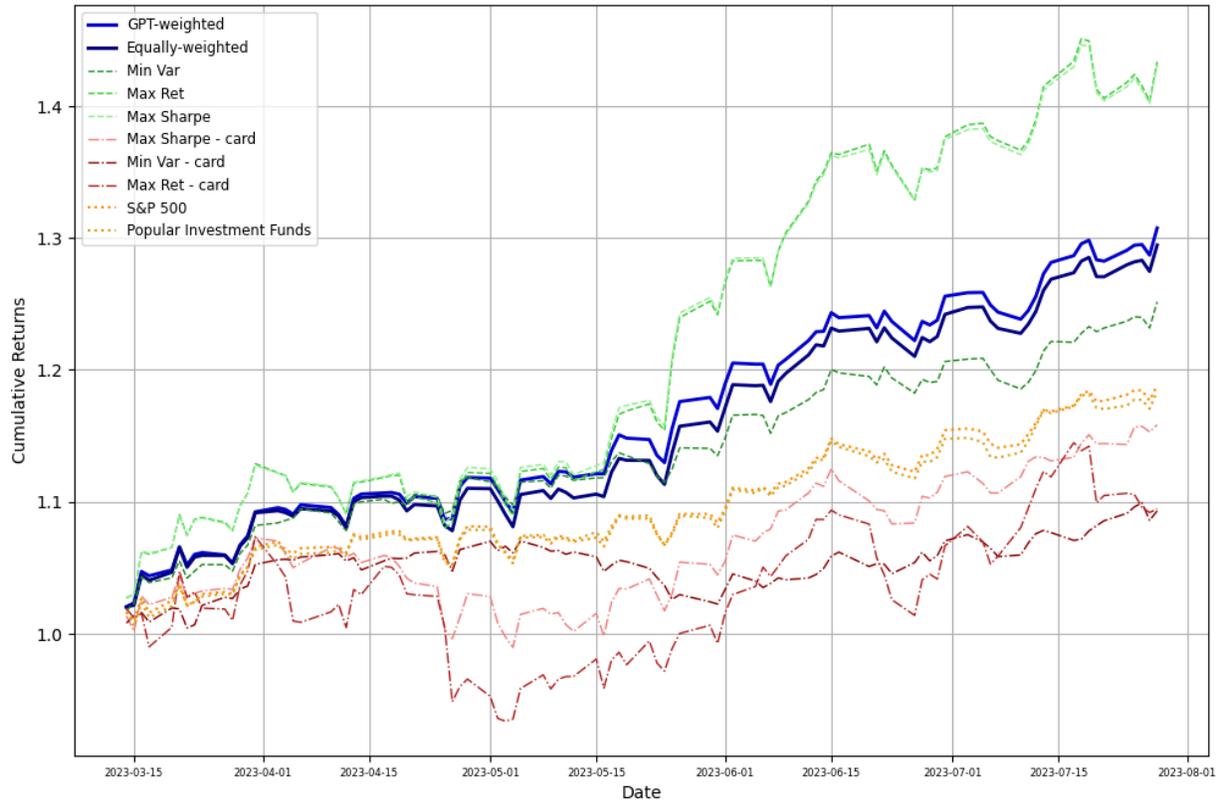

**Fig. 10** Cumulative returns of portfolios with 15 assets and benchmarks out-of-sample (14 March 2023 – 31 July 2023)

| | Cumulative Returns | Expected Return | Volatility of Return | Max Drawdown | Sharpe Ratio | VaR 99% of Return |
|---|---|---|---|---|---|---|
| **GPT-weighted** | 130.77% | 1.26% | 1.26% | -0.96% | 4.82 | -1.32% |
| **Equally-weighted** | 129.48% | 1.21% | 1.31% | -0.85% | 4.65 | -1.36% |
| **Min Var** | 125.18% | 1.05% | 1.07% | -1.18% | 4.74 | -1.19% |
| **Max Ret** | 143.39% | 1.71% | 2.21% | -2.24% | 4.58 | -1.98% |
| **Max Sharpe** | 143.17% | 1.70% | 2.15% | -1.77% | 4.64 | -1.90% |
| **Max Sharpe - card** | 115.83% | 0.69% | 1.94% | -6.61% | 2.24 | -2.05% |
| **Min Var - card** | 109.30% | 0.42% | 1.40% | -3.81% | 2.28 | -1.23% |
| **Max Ret - card** | 109.46% | 0.42% | 3.56% | -10.73% | 0.89 | -3.89% |
| **S&P 500** | 118.84% | 0.79% | 1.31% | -1.39% | 3.57 | -1.59% |
| **Dow Jones** | 111.44% | 0.50% | 1.46% | -2.95% | 2.40 | -1.22% |
| **NASDAQ** | 127.95% | 1.14% | 1.58% | -1.44% | 3.92 | -1.98% |
| **Popular Investment Funds** | 118.38% | 0.77% | 1.40% | -1.65% | 3.44 | -1.43% |

**Table 3** Evaluation metrics (weekly) for portfolios with 15 stocks and benchmarks from 14 March 2023 to 31 July 2023

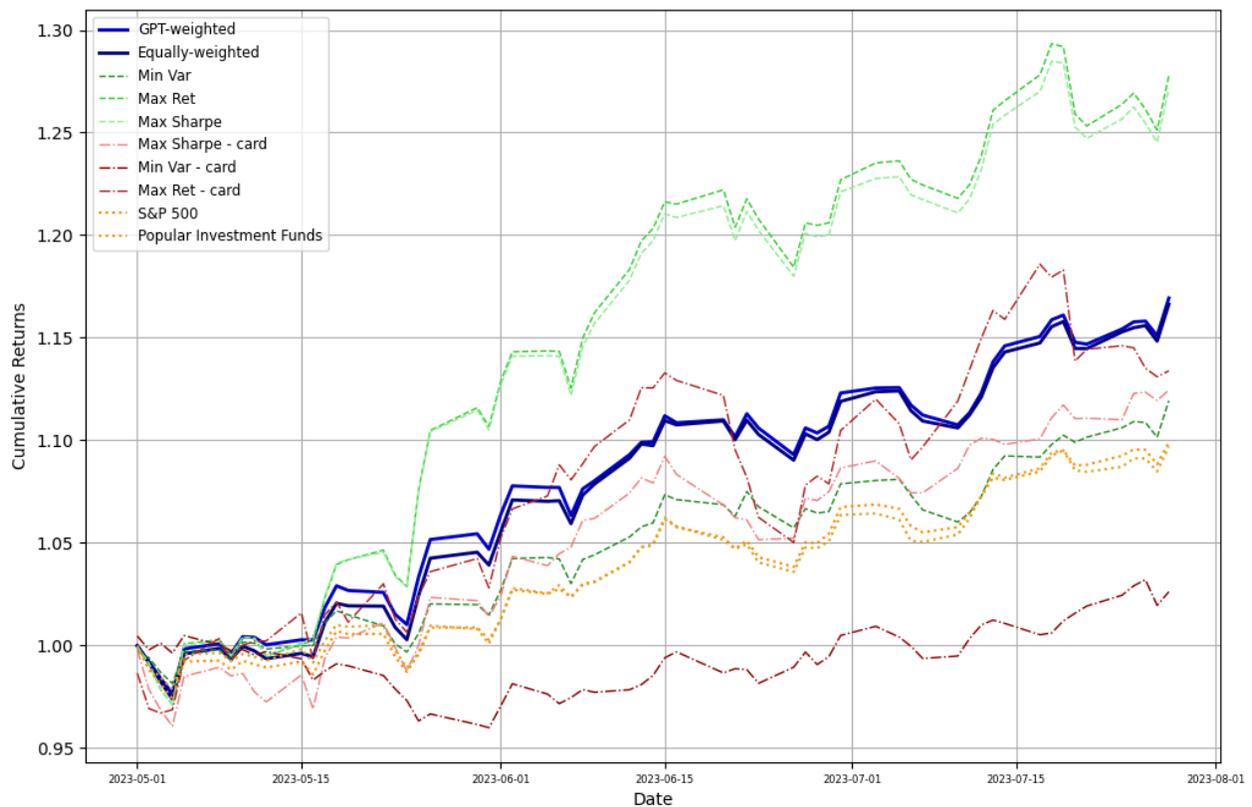

**Fig. 11** Cumulative returns of portfolios with 15 assets and benchmarks out-of-sample (1 May 2023 – 31 July 2023)

|  | Cumulative Returns | Expected Return | Volatility of Return | Max Drawdown | Sharpe Ratio | VaR 99% of Return |
|---|---|---|---|---|---|---|
| GPT-weighted | 116.93% | 1.22% | 1.38% | -0.96% | 4.77 | -1.21% |
| Equally-weighted | 116.63% | 1.20% | 1.42% | -0.85% | 4.72 | -1.13% |
| Min Var | 111.93% | 0.90% | 1.15% | -1.18% | 4.27 | -1.04% |
| Max Ret | 127.78% | 1.93% | 2.33% | -0.97% | 5.07 | -2.18% |
| Max Sharpe | 127.14% | 1.89% | 2.29% | -0.90% | 5.05 | -2.09% |
| Max Sharpe - card | 112.45% | 0.94% | 1.95% | -2.94% | 3.19 | -1.80% |
| Min Var - card | 102.61% | 0.17% | 1.48% | -3.81% | 1.00 | -1.12% |
| Max Ret - card | 113.38% | 1.12% | 2.94% | -5.91% | 2.76 | -2.92% |
| S&P 500 | 109.90% | 0.74% | 1.37% | -1.39% | 3.63 | -1.14% |
| Dow Jones | 103.99% | 0.32% | 1.55% | -1.96% | 1.62 | -1.07% |
| NASDAQ | 117.09% | 1.24% | 1.67% | -1.44% | 4.44 | -1.60% |
| Popular Investment Funds | 109.70% | 0.73% | 1.47% | -1.66% | 3.45 | -1.13% |

**Table 4** Evaluation metrics (weekly) for portfolios with 15 stocks and benchmarks from 1 May 2023 to 31 July 2023

Across different out-of-sample time periods and portfolios of 15, 30, and 45 stocks, we can observe some distinct patterns. Consistently, the "Max Ret" and "Max Sharpe" strategies were observed to provide the highest cumulative returns. However, these higher returns were often paired with increased risk measures (volatility, max drawdown, value-at-risk), indicating a greater level of risk associated with these portfolios.

In addition, "Max Ret" and "Max Sharpe" strategies have the highest Sharpe ratios indicating that those portfolios have the best out-of-sample performance per unit of risk (volatility). On the other hand, the "Min Var" strategy generally displayed the lowest weekly volatility, making it the least risky option despite often producing lower returns. It was also noted that the "Max Ret - card" strategy, while showing high volatility, typically failed to produce high returns. This trend was especially evident in the 30 and 45 stock portfolios. The "GPT-weighted" and "equally-weighted" portfolios provided solid performance, generally falling between the high returns-high risk "Max Ret" and "Max Sharpe" strategies, and the low risk-low return "Min Var" strategy.

In essence, while "Max Ret" and "Max Sharpe" strategies offered a potential for high returns and had superior Sharpe ratios, those also carried substantial risk. Investors seeking stability over high returns would likely be better served by the "Min Var" strategy. Those seeking a balance between risks and returns might consider "GPT-weighted" or "equally-weighted" portfolios.

Figure 12 compares the three top-performing portfolios ("GPT-weighted", "Max Ret" and "Max Sharpe") of the three universes. We observe that the 15 stock portfolios perform the best with respect to cumulative return, followed by the 30 stock portfolios and the relatively worse performance is shown by the 45 stock portfolios. This leads us to conclude that the lesser the number of stocks in the portfolio (out of sizes 15, 30 and 45), the better the cumulative return. The 15 stock portfolios demonstrated superior performance and stability, possibly due to a well-selected group of stocks that have stronger fundamentals. On the other hand, the 45 stock portfolios, while more diversified, may have been affected by lower-performing assets, resulting in relatively worse overall performance.

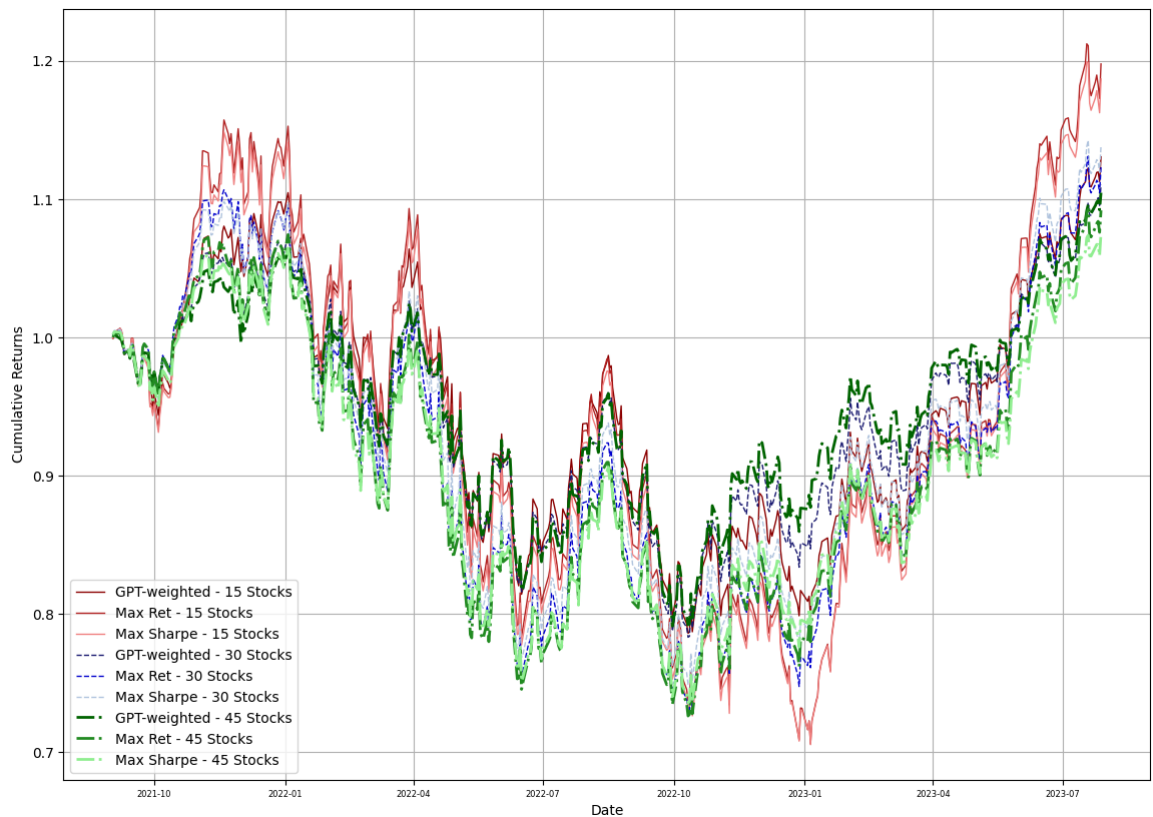

**Fig. 12** Cumulative returns for "GPT-weighted", "Max Ret" and "Max Sharpe" portfolios with 15, 30 and 45 stocks

In analyzing the overall performance of the various strategies, it becomes clear that portfolios on the Markowitz mean-variance efficient frontier, specifically derived from the trading universes selected by ChatGPT, consistently outperform others for all out-of-sample time periods. These portfolios, optimized for maximum returns or maximum risk-adjusted returns, were notable for their superior performance, exhibiting a fine balance between risk and reward.

In contrast, the ChatGPT strategies, both "equally-weighted" and "GPT-weighted", demonstrated less optimal outcomes. This may suggest that while the AI has capabilities in selecting profitable stocks, its ability to optimally distribute portfolio weights could benefit from further refinement. Interestingly, the cardinality-constrained optimized portfolios fell short of expectations. Despite their intent to enhance diversification and improve stability by limiting the number of assets, those did not yield high returns out-of-sample, suggesting potential overfitting to the in-sample data. Robust parameter estimation or applying robust mean-variance optimization may improve out-of-sample risk-return profiles of portfolios produced by cardinality-constrained optimization.

Despite this, it's remarkable that the majority of implemented strategies outperformed traditional benchmarks, such as S&P500, underscoring the potential of integrating AI capabilities like ChatGPT into investment strategies design. Moreover, strategies that augmented performance of ChatGPT's selected portfolios through the application of optimization techniques, saw a significant improvement in risk-return profiles. This showcases the value of supplementing AI capabilities with established financial theories and techniques.

## 5 Conclusions

Utilizing ChatGPT as a "standalone" investment advisor may not necessarily be a good idea due to a number of limitations. First, the generative models "hallucinate" and hence outputs produced by those require rigorous validation. It makes an implementation of robust validation checks on its recommendations a requirement for practical use. At the minimum, obtaining reliable investment universes involves making multiple requests to ChatGPT, assessing the results, and using a majority voting system to finalize a universe of stocks. Second, it is imperative to ensure that the assets recommended are indeed part of an index that we requested, e.g., S&P500. Additionally, other less obvious checks and validations of ChatGPT output may need to be performed.

On the other hand, it is worth noting that ChatGPT basically leverages a combined sentiment from the period in which it was trained, which may not always align with the most recent market data, at least for now. However, when ChatGPT's output is combined with traditional quantitative models, there is a significant potential to improve risk-reward profiles of investment portfolios. By utilizing investment universes generated by ChatGPT and performing portfolio optimization on top of those, we can obtain better performing investment portfolios than those obtained exclusively using ChatGPT or portfolio optimization algorithms without ChatGPT.

In summary, our findings lead to two key conclusions. Firstly, the superior performance of ChatGPT portfolios compared to the cardinality-constrained portfolios reinforces the notion that ChatGPT excels in "stock picking". Secondly, when combined with rigorous validation and traditional quantitative financial modeling, such as portfolio optimization, ChatGPT proves to be a valuable tool for investment. These hybrid solutions offer the potential for more efficient and effective approaches to practical investing, presenting themselves as superior alternatives to current robo-advising strategies.

**Acknowledgements** The authors' research was partially supported by the Mitacs Globalink Research Internship program and by SS&C Technologies (SS&C Algorithmics). We would like to thank Rafael Mendoza-Arriaga from SS&C Algorithmics for his helpful suggestions.

# Appendix A

Figures 13, 14, and 15 illustrate out-of-sample cumulative returns for the 30 stock universe, and Figures 16, 17 and 18 for the 45 stock universe during period 1, period 2, and period 3, respectively. In Tables 5, 6, and 7, we compare out-of-sample performance and risk metrics of various portfolios within the 30 stock universe, and in Tables 8, 9 and 10 comparison for the 45 stock universe is provided.

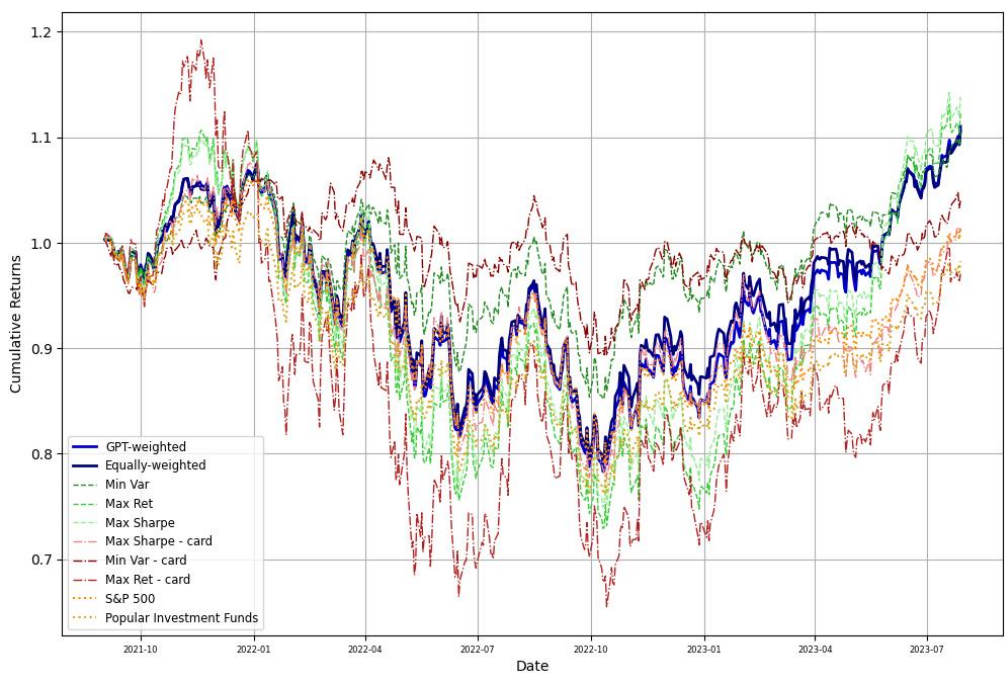

**Fig. 13** Cumulative returns of portfolios with 30 assets and benchmarks out-of-sample (1 September 2021 – 31 July 2023)

|  | Cumulative Returns | Expected Return | Volatility of Return | Max Drawdown | Sharpe Ratio | VaR 99% of Return |
|---|---|---|---|---|---|---|
| GPT-weighted | 111.02% | 0.14% | 2.78% | -25.71% | 0.36 | -3.63% |
| Equally-weighted | 110.63% | 0.13% | 2.68% | -24.92% | 0.36 | -3.52% |
| Min Var | 110.38% | 0.12% | 2.30% | -19.52% | 0.38 | -3.03% |
| Max Ret | 112.29% | 0.17% | 3.44% | -33.85% | 0.36 | -4.12% |
| Max Sharpe | 113.83% | 0.18% | 3.30% | -31.56% | 0.40 | -4.01% |
| Max Sharpe - card | 101.41% | 0.05% | 3.06% | -27.53% | 0.12 | -3.62% |
| Min Var - card | 103.97% | 0.06% | 2.11% | -17.16% | 0.20 | -2.40% |
| Max Ret - card | 97.03% | 0.08% | 4.82% | -45.06% | 0.13 | -5.43% |
| S&P 500 | 101.29% | 0.04% | 2.62% | -24.82% | 0.13 | -3.42% |
| Dow Jones | 100.42% | 0.03% | 2.32% | -20.95% | 0.09 | -2.80% |
| NASDAQ | 93.52% | -0.01% | 3.31% | -35.72% | -0.00 | -4.20% |
| Popular Investment Funds | 98.25% | 0.01% | 2.55% | -26.00% | 0.04 | -3.32% |

**Table 5** Evaluation metrics (weekly) for portfolios with 30 stocks and benchmarks from 1 Sept 2021 to 31 July 2023

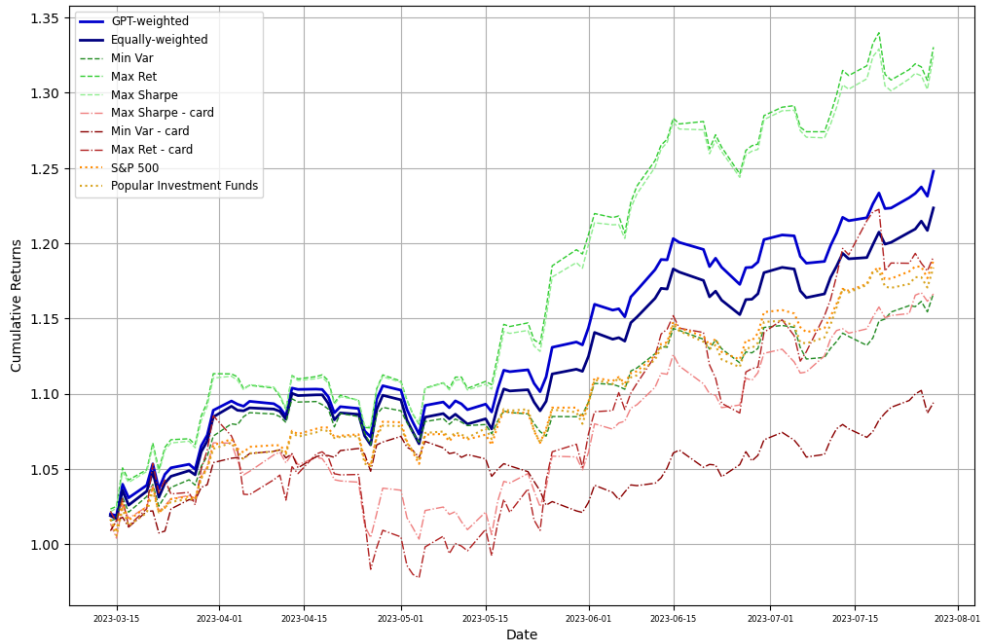

**Fig. 14** Cumulative returns of portfolios with 30 assets and benchmarks out-of-sample (14 March 2023 – 31 July 2023)

|  | Cumulative Returns | Expected Return | Volatility of Return | Max Drawdown | Sharpe Ratio | VaR 99% of Return |
|---|---|---|---|---|---|---|
| GPT-weighted | 124.79% | 1.02% | 1.46% | -1.44% | 4.29 | -1.38% |
| Equally-weighted | 122.35% | 0.93% | 1.47% | -1.74% | 4.02 | -1.33% |
| Min Var | 116.65% | 0.70% | 1.25% | -1.83% | 3.57 | -1.19% |
| Max Ret | 133.04% | 1.34% | 1.77% | -1.28% | 4.49 | -1.69% |
| Max Sharpe | 132.45% | 1.32% | 1.71% | -1.17% | 4.61 | -1.65% |
| Max Sharpe - card | 116.63% | 0.72% | 1.81% | -5.41% | 2.59 | -1.84% |
| Min Var - card | 109.40% | 0.42% | 1.39% | -3.74% | 2.27 | -1.37% |
| Max Ret - card | 119.09% | 0.81% | 3.00% | -8.27% | 1.95 | -3.27% |
| S&P 500 | 118.84% | 0.79% | 1.31% | -1.39% | 3.57 | -1.59% |
| Dow Jones | 111.44% | 0.50% | 1.46% | -2.95% | 2.40 | -1.22% |
| NASDAQ | 127.95% | 1.14% | 1.58% | -1.44% | 3.92 | -1.98% |
| Popular Investment Funds | 118.38% | 0.77% | 1.40% | -1.65% | 3.44 | -1.43% |

**Table 6** Evaluation metrics (weekly) for portfolios with 30 stocks and benchmarks from 14 March 2023 to 31 July 2023

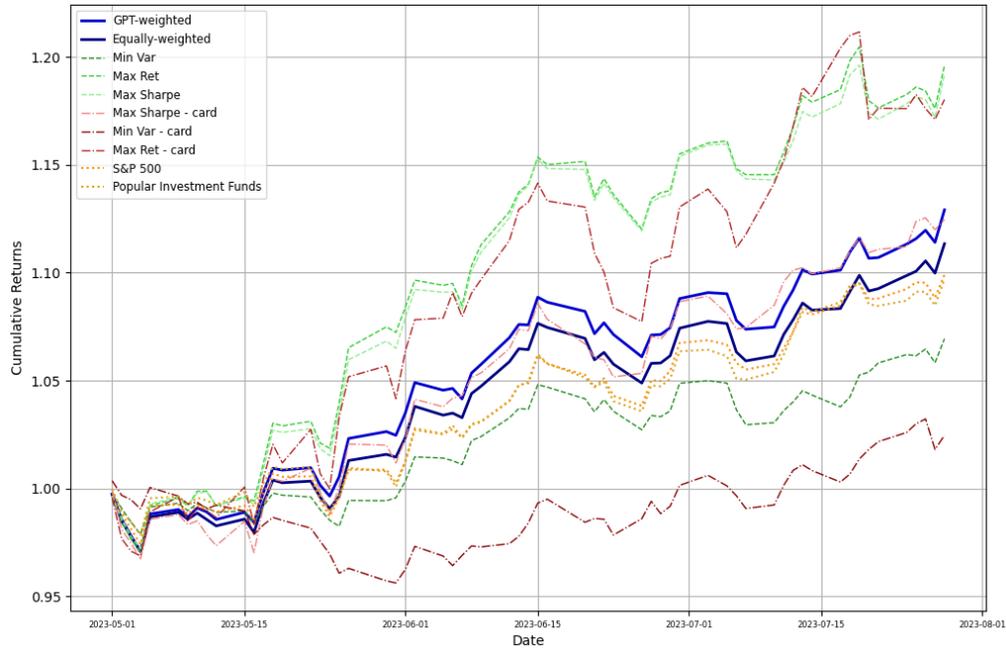

**Fig. 15** Cumulative returns of portfolios with 30 assets and benchmarks out-of-sample (1 May 2023 – 31 July 2023)

|  | Cumulative Returns | Expected Return | Volatility of Return | Max Drawdown | Sharpe Ratio | VaR 99% of Return |
|---|---|---|---|---|---|---|
| GPT-weighted | 112.90% | 0.97% | 1.50% | -1.38% | 4.43 | -1.15% |
| Equally-weighted | 111.33% | 0.86% | 1.50% | -1.58% | 4.09 | -1.23% |
| Min Var | 106.93% | 0.54% | 1.24% | -1.83% | 3.09 | -0.97% |
| Max Ret | 119.59% | 1.43% | 1.78% | -1.21% | 4.98 | -1.68% |
| Max Sharpe | 119.19% | 1.40% | 1.73% | -1.17% | 5.05 | -1.56% |
| Max Sharpe - card | 112.45% | 0.93% | 1.86% | -2.48% | 3.57 | -1.59% |
| Min Var - card | 102.45% | 0.17% | 1.49% | -3.74% | 0.97 | -1.21% |
| Max Ret - card | 118.01% | 1.35% | 2.75% | -4.37% | 3.44 | -2.54% |
| S&P 500 | 109.90% | 0.74% | 1.37% | -1.39% | 3.63 | -1.14% |
| Dow Jones | 103.99% | 0.32% | 1.55% | -1.96% | 1.62 | -1.07% |
| NASDAQ | 117.09% | 1.24% | 1.67% | -1.44% | 4.44 | -1.60% |
| Popular Investment Funds | 109.70% | 0.73% | 1.47% | -1.66% | 3.45 | -1.13% |

**Table 7** Evaluation metrics (weekly) for portfolios with 30 stocks and benchmarks from 1 May 2023 to 31 July 2023

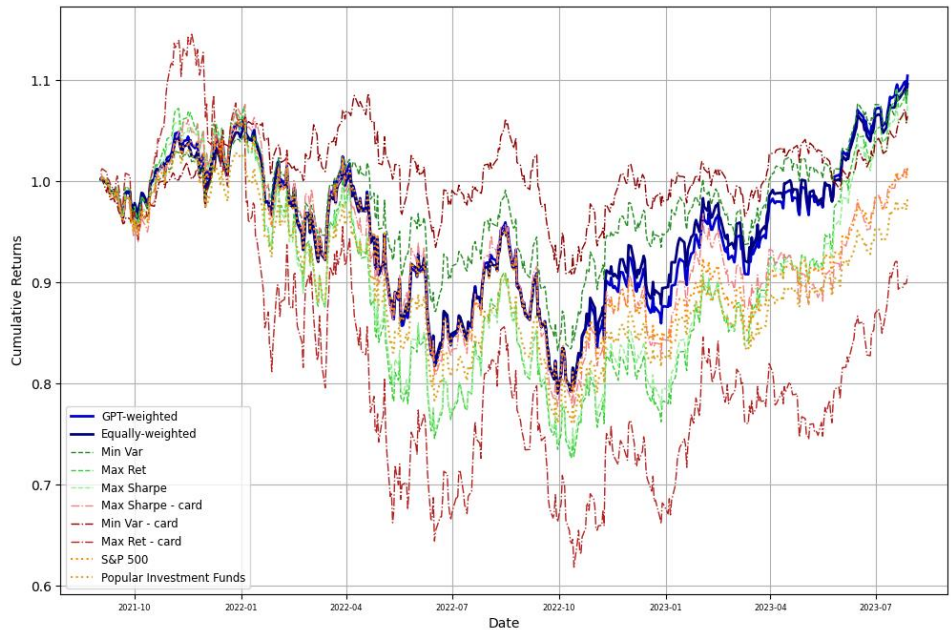

**Fig. 16** Cumulative returns of portfolios with 45 assets and benchmarks out-of-sample (1 September 2021 – 31 July 2023)

| | Cumulative Returns | Expected Return | Volatility of Return | Max Drawdown | Sharpe Ratio | VaR 99% of Return |
|---|---|---|---|---|---|---|
| **GPT-weighted** | 110.45% | 0.13% | 2.69% | -24.72% | 0.36 | -3.57% |
| **Equally-weighted** | 109.65% | 0.12% | 2.59% | -24.28% | 0.34 | -3.43% |
| **Min Var** | 109.18% | 0.11% | 2.27% | -20.16% | 0.35 | -2.92% |
| **Max Ret** | 109.18% | 0.14% | 3.27% | -31.99% | 0.31 | -4.06% |
| **Max Sharpe** | 107.57% | 0.12% | 3.09% | -30.37% | 0.28 | -3.83% |
| **Max Sharpe - card** | 101.32% | 0.05% | 3.04% | -27.13% | 0.12 | -3.51% |
| **Min Var - card** | 106.48% | 0.08% | 2.09% | -16.25% | 0.29 | -2.43% |
| **Max Ret - card** | 90.63% | 0.00% | 4.73% | -46.10% | 0.02 | -5.37% |
| **S&P 500** | 101.29% | 0.04% | 2.62% | -24.82% | 0.13 | -3.42% |
| **Dow Jones** | 100.42% | 0.03% | 2.32% | -20.95% | 0.09 | -2.80% |
| **NASDAQ** | 93.52% | -0.01% | 3.31% | -35.72% | -0.00 | -4.20% |
| **Popular Investment Funds** | 98.25% | 0.01% | 2.55% | -26.00% | 0.04 | -3.32% |

**Table 8** Evaluation metrics (weekly) for portfolios with 45 stocks and benchmarks from 1 September 2021 to 31 July 2023

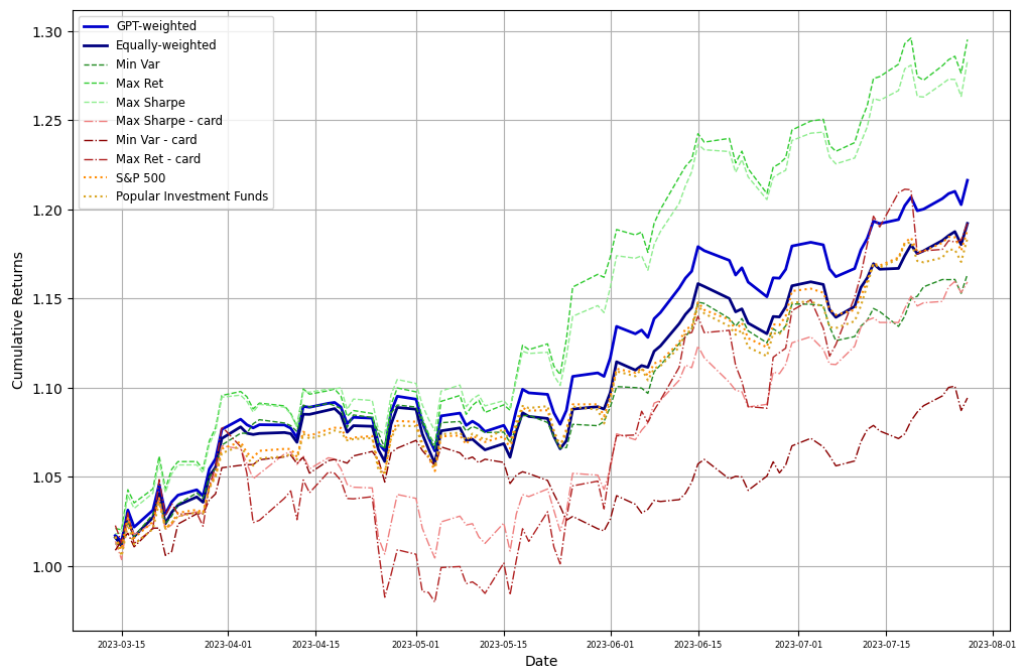

**Fig. 17** Cumulative returns of portfolios with 45 assets and benchmarks out-of-sample (14 March 2023 – 31 July 2023)

|  | Cumulative Returns | Expected Return | Volatility of Return | Max Drawdown | Sharpe Ratio | VaR 99% of Return |
|---|---|---|---|---|---|---|
| **GPT-weighted** | 121.63% | 0.91% | 1.46% | -1.81% | 3.97 | -1.33% |
| **Equally-weighted** | 119.21% | 0.81% | 1.51% | -2.18% | 3.64 | -1.30% |
| **Min Var** | 116.36% | 0.71% | 1.33% | -1.82% | 3.63 | -1.07% |
| **Max Ret** | 129.52% | 1.21% | 1.66% | -1.25% | 4.32 | -1.67% |
| **Max Sharpe** | 128.33% | 1.18% | 1.58% | -1.34% | 4.46 | -1.53% |
| **Max Sharpe - card** | 115.93% | 0.69% | 1.76% | -5.12% | 2.49 | -1.77% |
| **Min Var - card** | 109.43% | 0.42% | 1.39% | -3.65% | 2.28 | -1.26% |
| **Max Ret - card** | 119.21% | 0.81% | 2.98% | -8.68% | 2.02 | -2.87% |
| **S&P 500** | 118.84% | 0.79% | 1.31% | -1.39% | 3.57 | -1.59% |
| **Dow Jones** | 111.44% | 0.50% | 1.46% | -2.95% | 2.40 | -1.22% |
| **NASDAQ** | 127.95% | 1.14% | 1.58% | -1.44% | 3.92 | -1.98% |
| **Popular Investment Funds** | 118.38% | 0.77% | 1.40% | -1.65% | 3.44 | -1.43% |

**Table 9** Evaluation metrics (weekly) for portfolios with 45 stocks and benchmarks from 14 March 2023 to 31 July 2023

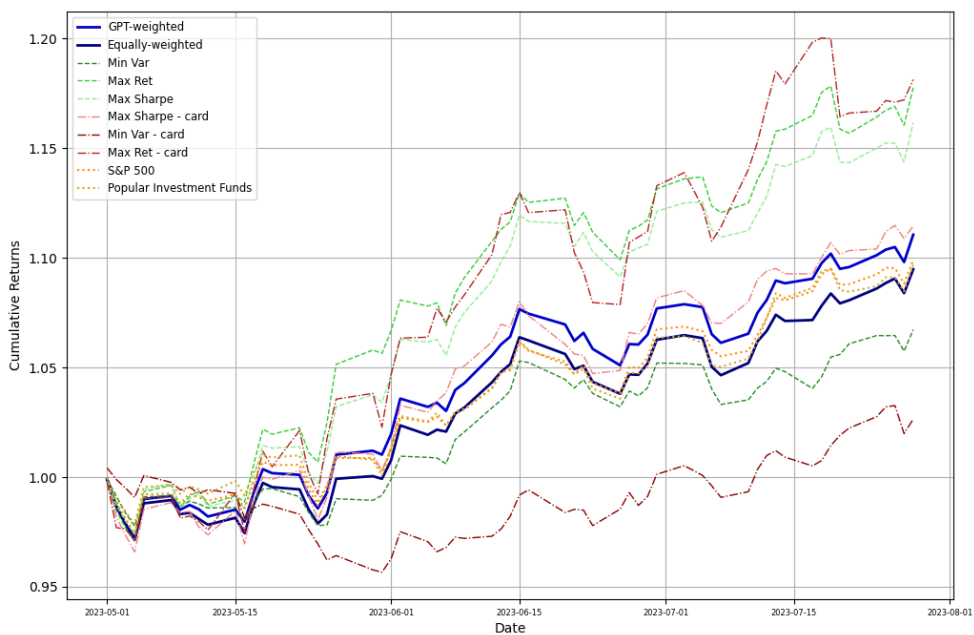

**Fig. 18** Cumulative returns of portfolios with 45 assets and benchmarks out-of-sample (1 May 2023 – 31 July 2023)

| | Cumulative Returns | Expected Return | Volatility of Return | Max Drawdown | Sharpe Ratio | VaR 99% of Return |
|---|---|---|---|---|---|---|
| GPT-weighted | 111.06% | 0.83% | 1.57% | -1.50% | 3.89 | -1.16% |
| Equally-weighted | 109.47% | 0.72% | 1.61% | -1.78% | 3.45 | -1.25% |
| Min Var | 106.72% | 0.52% | 1.41% | -1.82% | 2.88 | -0.89% |
| Max Ret | 117.76% | 1.30% | 1.78% | -1.21% | 4.73 | -1.47% |
| Max Sharpe | 116.17% | 1.19% | 1.70% | -1.23% | 4.68 | -1.24% |
| Max Sharpe - card | 111.45% | 0.87% | 1.79% | -2.47% | 3.36 | -1.56% |
| Min Var - card | 102.64% | 0.18% | 1.51% | -3.65% | 1.03 | -1.17% |
| Max Ret - card | 118.13% | 1.35% | 2.86% | -3.66% | 3.54 | -2.43% |
| S&P 500 | 109.90% | 0.74% | 1.37% | -1.39% | 3.63 | -1.14% |
| Dow Jones | 103.99% | 0.32% | 1.55% | -1.96% | 1.62 | -1.07% |
| NASDAQ | 117.09% | 1.24% | 1.67% | -1.44% | 4.44 | -1.60% |
| Popular Investment Funds | 109.70% | 0.73% | 1.47% | -1.66% | 3.45 | -1.13% |

**Table 10** Evaluation metrics (weekly) for portfolios with 45 stocks and benchmarks from 1 May 2023 to 31 July 2023

# Appendix B

The following prompts were used to extract trading universes and "GPT-weighted" portfolios of stocks from GPT-4:

- Prompt 1: "Using a range of investing principles taken from leading funds, create a theoretical fund comprising of at least X stocks (mention their tickers) from the S&P500 with the goal to outperform the S&P500 index" (X is 15, 30 or 45 for the three cases);
- Prompt 2: "Extract only the ticker symbols of the stocks comprising the fund from the previous response: - '{input}'. In your response to this prompt, list only the ticker symbols separated by spaces" (here the output response of prompt 1 was used in place of {input});

- Prompt 3: "Assume you're designing a theoretical model portfolio from these S&P500 stocks: {input}. Provide a hypothetical example of how you might distribute the weightage of these stocks (normalized, i.e., weights should add up to 1.00) in the portfolio to potentially outperform the S&P500 index. Also mention the underlying strategy or logic which you used to assign these weights" (here the output response of prompt 2 was used in place of {input});
- Prompt 4: "Extract tickers of stocks and corresponding weights as a single comma ',' separated string, with the weights expressed as floats: '{input}' " (here the output response of prompt 3 was used in place of {input}).

Note 1: Prompts 1 and 2 were used together and this was iterated 30 times for each of the three cases.

Note 2: In some cases, the weights assigned by ChatGPT may not add up to exactly 1 (it was 1.05 in the case of 15 stocks for our experiment, see Figure 20). In such cases, we normalized weights such that those add up to exactly 1. This ensures that the relative weight distribution remains the same.

Figure 19 shows an output response of GPT-4 model in the ChatGPT user interface for prompt 1 (X = 15).

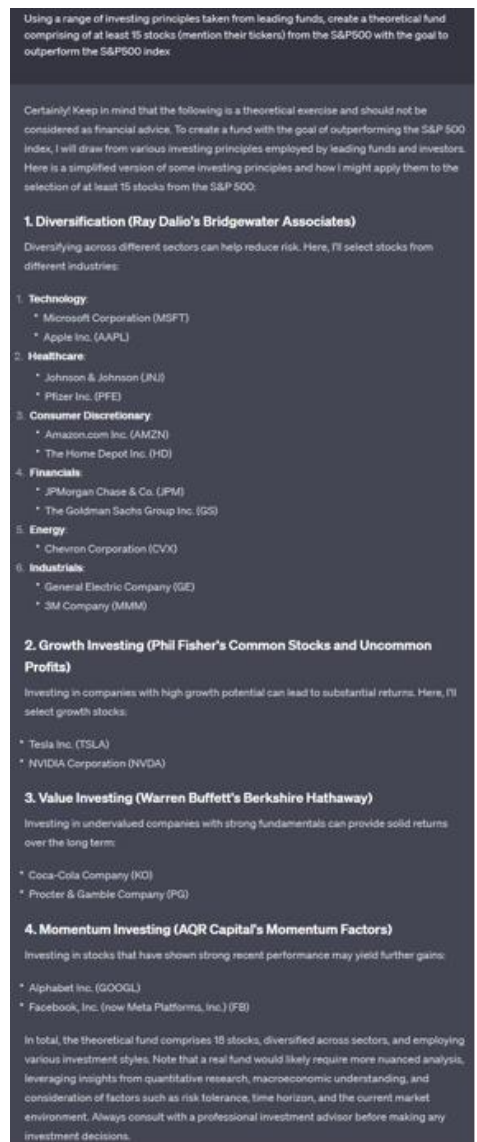

**Fig. 19** An output response of ChatGPT (GPT-4) for prompt 1 with X = 15

Figure 20 shows an output response of GPT-4 model in the ChatGPT user interface for prompt 3 (for the stocks obtained in the output seen in Figure 19).

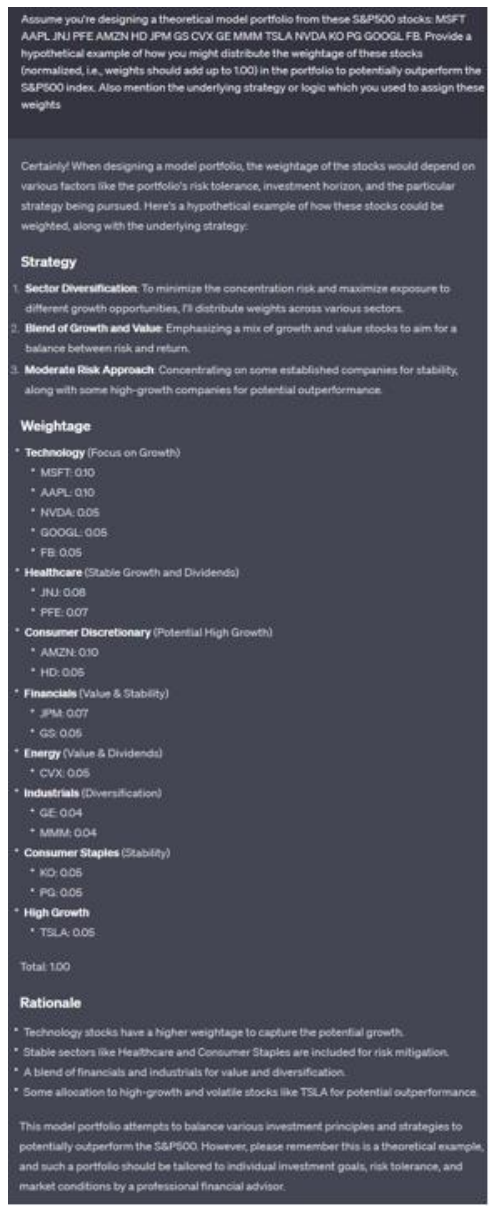

**Fig. 20** An output response of ChatGPT (GPT-4) for prompt 3 with X = 15

Note 3: The outputs shown in Figure 19 and Figure 20 represent responses obtained through the GPT-4 user interface for the corresponding prompts used in the experiment. These outputs may differ from the responses obtained through the GPT-4 API, which were used in the actual experiment. The difference is due to the probabilistic nature of the GPT-4 model, which can generate varying responses to the same prompt between different calls or interfaces.